# Tri-Compress: A Cascaded Data Compression Framework for Smart Electricity Distribution Systems

Syed Muhammad Atif[1*], Anees Ahmed[2], Sameer Qazi[3]

[1,2]Graduate School of Science and Engineering, PAF Karachi Institute of Economics and Technology,
Karachi, 75190, Pakistan

[3] College of Engineering, PAF Karachi Institute of Economics and Technology,
Karachi, 75190, Pakistan

syed.muhammad.atif@gmail.com[1], aneesemail@yahoo.com[2], sameer.qazi@pafkiet.edu.pk[3]

* Corresponding author

**Abstract**: Modern smart distribution system requires storage, transmission and processing of big data generated by sensors installed in electric meters. On one hand, this data is essentially required for intelligent decision making by smart grid but on the other hand storage, transmission and processing of that huge amount of data is also a challenge. Present approaches to compress this information have only relied on the traditional matrix decomposition techniques benefitting from low number of principal components to represent the entire data. This paper proposes a cascaded data compression technique that blends three different methods in order to achieve high compression rate for efficient storage and transmission. In the first and second stages, two lossy data compression techniques are used, namely Singular Value Decomposition (SVD) and Normalization; Third stage achieves further compression by using the technique of Sparsity Encoding (SE) which is a lossless compression technique but only having appreciable benefits for sparse data sets. Our simulation results show that the combined use of the 3 techniques achieves data compression ratio to be 15% higher than state of the art SVD for small, sparse datasets and up to 28% higher in large, non-sparse datasets with acceptable Mean Absolute Error (MAE).

**Keywords:** Singular Value Decomposition, Sparse Matrix Representation, Smart Grid, Data Compression, Big Data,

## I. INTRODUCTION

Internet of Things (IoT) is a fast moving technology referring to a ubiquitous network of 'always connected' smart gadgets that can remotely perform monitoring and reporting of physical parameters of their surroundings to a remote entity through a network. The remote entity may make rapid decisions based on artificially intelligent algorithms or manual intervention by humans. Power distribution systems are also taking advantage of this concept through smart meters. Smart meters are rapidly replacing the conventional meters that allows power distributions companies to get insight of user power usage behavior and make their decision accordingly. Power grids are not only vital in distribution system but also very sensitive to fluctuations in demands. Sudden increase in demand leads to tripping of grid stations. Unfortunately, this local failure has cascading effects under certain circumstances leading to a catastrophe such as New York power outage 2003 [1-3]. Those catastrophic events urges us to design and deploy such power distribution system that are smart and intelligent enough to quickly take their decisions both reactively and proactively without any human intervention. As a result, a new era of smart power distribution systems begins. They are equipped with smart metering system that continuously provide power usage of their users. The system then utilizes this data for operations like monitoring, analysis and control. However, smart meter and other similar equipment continuously generate huge amount of data at constant rate that bring the challenges in terms of their transmission and storage. This paper focus the said problem and proposed a new solution named tri compress. As the name suggest this solution is the blend of three different methods precisely, Singular Value Decomposition (SVD), normalization and value-index sparse matrix representation. In nature, the proposed technique is a lossy data compression technique. But, small degree of information loss is affordable as this data from metering devices is mostly used for monitoring and planning purposes that require coarse grained information.

This paper is organized as follows. Section I has presented the introduction. Section II will give related work. Section III, the core of this paper, will present the proposed idea and technique for data compression. Section IV will experimentally evaluate the newly proposed idea. Section V will provide the obtained results of simulation. Finally, section VI will conclude the paper.

## II. RELATED WORK

With the advancements in smart distributions systems and integration of IoT, the amount of data available of storage, transmission and processing will become gigantic. This challenge draws attention of many researchers towards the development of data compression technique that specially cater needs of smart distribution systems [3-10]. Brindha and D. Sundararajan [7] use discrete wavelet techniques for compression of data. They use a bi-orthogonal 5/3 spline filter for compression and were able to achieve a compression ratio of up to 8:1. Phasor measurement unit, a well-known image compression technique, is employed by Klump et al. [9] for compression of smart grid data that result in obtaining the best compression ratio of 14.35:1. Kapischet al. presented the concept of gapless waveform recording and Novelty Detection Concept in [16] in order to attain compression. However, there is still a need of higher compression ratio due to large amount of data which is specifically addressed in this paper.

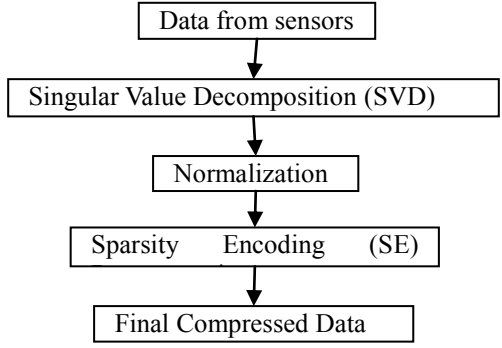

**Fig. 1 Data Compression Framework**

The nature of data obtained from smart grid system allow us to store it in the form matrix as the data coming from different sensor belong to measurements taken for the same device at several different time instances. This format of dataset representation is the most appropriate for SVD, a widely used technique in the field of image compression and many others [11-15]. This paper utilizes SVD for compression of data of smart grid systems. The reason is that it provides us a good tradeoff between information loss and degree of achieved compression. It is possible in case of smart grid data as it generally used by applications for monitoring and planning purposes that do not require high precision data.

### III. PROPOSED SOLUTION

In this section, we explain our tri-stage cascaded data compression framework via singular value decomposition (SVD), normalization and sparse matrix representation (see Fig. 1.). The subsection A explains how SVD will be utilized in our data compression framework model to achieve data compression. Subsection B will elaborates the second stage of our model i.e. normalization for efficient representation of data obtain after apply SVD in the first stage whereas subsection C provides detail regarding the third stage of our model that exploits sparse matrix representation for compressed data obtained from second stage so that data can be stored or transmitted in its most compressed form.

### A. Singular value decomposition (SVD)

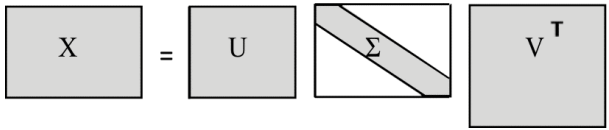

**Fig. 2 SVD of matrix X**

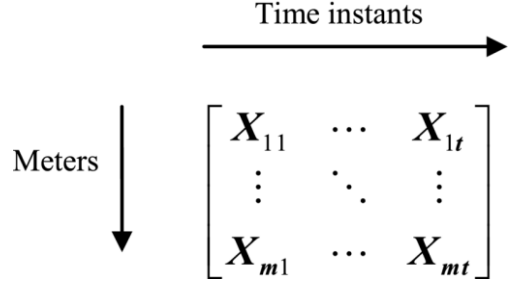

**Fig. 3 Representation of data as matrix X**

Let **X** be the data collected from different sensors at regular time intervals. This data **X** can be considers as a matrix of dimension $m$ by $t$ where $m$ is number of metering devices equipped with sensors and $t$ is the time stamps as depicted in Fig. 2. Singular value decomposition (SVD) will decompose this matrix **X** into three matrices **U**, **S** and **V** where **U** and **V** be the orthogonal unitary matrices of dimension $m$ by $m$ and $t$ by $t$ respectively where **S** be the diagonal matrix of dimension $m$ by $t$ that diagonal entries are arranged in descending order. Fig. 3 is giving a pictorial representation of this factorization. However, the effective dimension of **U**, **S** and **V** matrices are $m$ by $r$, $r$ by $r$ and $r$ by $t$ where $r$ is the rank of the matrix **X**.

In essence, SVD represents a matrix X as the sum of product of rank one matrices ordered in descending order of their respective Frobenius norm.

$$X \approx X_k = \sum_{i=1}^{k} \sigma_i u_i v_i^T \quad (1)$$

Since U and V are unitary matrices i.e. their column has unit L2 norm, so their corresponding singular values in the diagonal matrix S will represent the Frobenius norm of respective sum term. Fig. 4 is illustrating the typical decay (decrease) in the singular values in SVD. Hence, a matrix can be well approximated by the first few terms. Let **X** be approximated as $\mathbf{X_k}$ by the top k singular values then

$$X \approx X_k = \sum_{i=1}^{k} \sigma_i u_i v_i^T \quad (2)$$

In matrix notation

$$X \approx X_k = U_k \Sigma_k V_k^T \quad (3)$$

Where the dimension of $U_k, \Sigma_k$ and $V_k$ are $m$ by $k$, $k$ by $k$ and $k$ by $t$ respectively and $k \leq r$.

The storage or transmission capacity required for X without any manipulation is:

$$space(X) = m \times t \quad (4)$$

However, the storage or transmission capacity required for $X_k$, the approximation of $X$, is:

$$space(X_k) = space(U_k) + space(\Sigma_k) + space(V_k) \quad (5)$$

where,

$$space(U_k) = m \times k \quad (6)$$

$$space(\Sigma) = k \quad (7)$$

$$space(V_k) = t \times k \quad (8)$$

Hence,

$$space(X_k) = m \times k + k + t \times k = (m+1+t)k \quad (9)$$

Therefore, we can obtain compressed data in the form of $X_k$ i.e. $space(X_k) < space(X)$ provided that $m \times t/(m+1+t)t < 1$ or equivalently $k \ll r \leq m$.

**B. Normalization**

Data coming from sensors in electrical devices as in our case is mainly wattage, voltage and current usage i.e. it is numerical in nature. Such a data is typically stored or transmitted by computing devices in floating point format or mantissa exponent format, a widely used standard for this format is IEEE 754. However, blindly using this format is not good in our case because of reason that people in vicinity usually have similar rage of electrical device as well as similar electrical usage behavior. It leads us to the conclusion that the variation in obtained data will be at most order of two or so. It is an ideal condition where technique like normalization can be readily use for data compression.

Normalization is way of representing decimal point numerical data. It transforms the entire data, by simply multiply them suitable powers of their system radix, so their exponents become identical. This transformations generally leads to precision loss. However, this precision loss is negligible when variation in data is very low.

Normalization allows us to transmit or store the exponent part of data only once as it is now common. This leads to significant reduction in the size of that that need to be store or transmitted. For example there is 25% reduction in the size of data after normalization if the data is represented in single precision floating point IEEE 754 format as mantissa to exponent ratio in this format is 3:1.

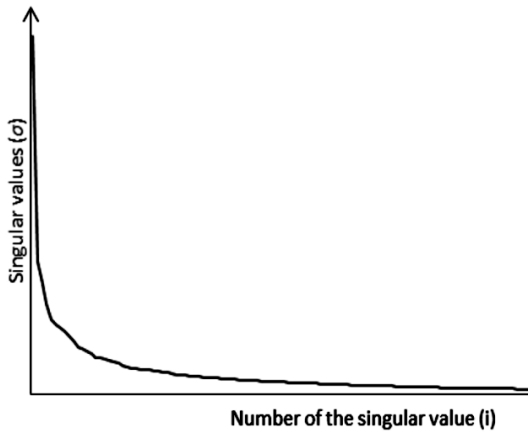

**Fig. 4 Typical decrease in ordered singular values**

**C. Sparsity Encoding (SE)**

Sparse matrix is a matrix where large number of entries are zeros. It is not a good idea to store (transmit) all the entries of such a matrix. Instead, one can store (transmit) a sparse matrix efficiently if it just store (transmit) only non-zero entries along with their indices (positions) in the matrix. It leads to reduction in size of data provided the amount of space required for zero entries is more than that required for index value representation of non-zero entries of the given sparse matrix. If the matrix X of our data is well sparse in nature then, $U_k$ and $V_k$ matrices obtained after application of SVD on this data will also be sparse matrices i.e. most of the entries in most of the column of $U_k$ (resp. most of the entries in most of the rows of $V_k$) will be zero. Thus, index value representation of $U_k$ and $V_k$ matrices will likely lead to significant data compression.

## IV. SIMULATION AND EXPERIMENTAL RESULTS

In this section, we will present results obtained from simulation conducted to check the validity of our proposed cascaded data compression framework specifically tailored for smart distribution (grid) system. We use three datasets. The first is a small, sparse dataset TU Darnstadt tracebase data set which depicts the energy usage of 80 devices that are switched on for a very limited time during a day or a week. This dataset is freely available from [17]. The second two Smart Grid datasets are medium and large (non-sparse) datasets also freely available from University of Massachusetts, UMass Smart Repository [18]. The first is Micro grid which has energy consumption information of 442 individual households and the second one is Apartments which has energy consumption information of 114 individual apartments. Table 1 summarizes the details of the 3 datasets used in this paper for evaluation of the efficiency of our scheme. Figure 5 shows the Eigen Spectrum of the 3 datasets on normalized scales for easy comparison. From Figure 5, we can observe the decay of the Eigen spectrum which is the square of the singular values, described earlier. The Eigen values are normalized with respect to the highest Eigen value in a given dataset. Similarly, the value of *k*, the number of principal components is normalized by the rank. This helps us to compare the dataset characteristics on one single figure. We observe that the UMass MicroGrid and Apartments datasets has a sharper exponential decay compared to the TU Damstadt dataset. This is due to the sparse nature of the TU Damstadt dataset as indicated in Table 1, which leads to a very high rank deficiency of the data matrix. This is because the five types of infrequently used appliances (total 80) considered in this dataset namely, namely 25 DVD player, 11 Subwoofer, 13 TV, 10 Hi-Fi Amplifier and 21 vacuum cleaners are switched on for a very limited time during the day. The other two datasets (UMass MicroGrid and UMass Apartments) do not suffer from this issue as they measure the aggregate energy consumption of 442 homes and 114 apartments respectively. Thus these datasets will tend to have better compression characteristics as will be evident in our next set of observations.

We design the MATLAB code to evaluate our methodology. As, our compression framework is lossy in nature so a metric is required to quantitatively measure the loss of information during compression. We use MAE as a metric to measure the loss of information during compression. It is defined as:

$$MAE = (\frac{1}{m \times t}) \sum_{i=1}^{m} \sum_{j=1}^{t} |X(i,j) - X_k(i,j)| \quad (10)$$

Figures 6-8, compares the storage requirements, Size reduction of compressed data of the original SVD and our proposed Tri-compress technique for all 3 datasets studied. For this study we assume pre-determined target compression ratios of 78, 39, 25, 19, 15, 9, 5 and 4. Then iteratively determine the number of principal components $k$ which will be required in each dataset to achieve the target compression ratio. Higher compression ratios will require smaller value of $k$ but will lead to high MAE as expected. You may observe that compression ratio decreases as $k$ "number of rank one matrices in SVD summation used for approximation" increases. This is expected as it leads to an increase in the dimensions of $U$, $S$ and $V$ matrices that in turn increases the size of file. The values of k in the UMass MicroGrid dataset is comparatively higher than the other two datasets considering it rank is highest 389 compared to 28 for TU Damstadt and 114 for UMass Apartments as shown in Table 1. Nevertheless, in all cases, higher value of $k$ means better approximation of given data or less loss of information. Therefore, there is a steady decline in MAE as the value of $k$ increases (Figure 8). It means that a tradeoff is required between compression ratio required and permissible loss of information while using our framework. One has to compromise on the precision of data if higher compression is required. Figure 9 again reveals the TU Damstadt dataset exhibits highest exponential reduction in the MAE as the number of principal components $k$ are added but it sets out with a much higher MAE value due to its abnormal sparsity leading to extreme rank deficiency and lower Eigen Spectrum decay (as was evident in Fig 5) compared to other two datasets UMass MicroGrid and UMass Apartments.

Normalization stage is always able to further compress the data obtain from SVD stage. However, its compression ability is less than both stage one and three for sparse datasets but more for large datasets with larger number of floating point data.

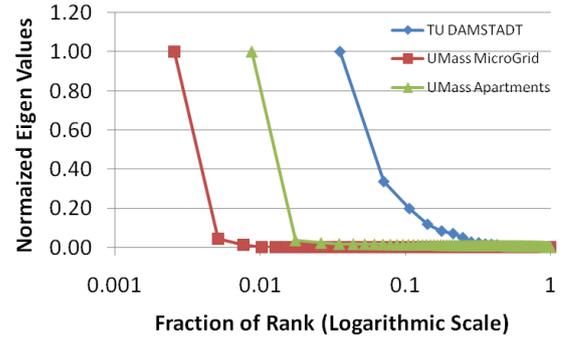

**Figure 5. Eigen Spectrum of the 3 datasets**

Sparse matrix representation always dramatically compresses the data if high level of sparsity present as in TU Damstadt dataset which is rank deficient as shown in Table 1. It will have limited or negligible benefits if dataset is non sparse such as UMass MicroGrid and UMass Apartments.

Figure 9 shows the MAE vs number of principal components normalized as a fraction of rank for easy comparison across 3 datasets for TriCompress scheme. It clearly shows that although error reduction is steep for TU Damstadt dataset due to extreme sparsity and resultant rank deficiency. MAE starts with a much smaller value for the UMass Apartments and UMass MicroGrid dataset.

**Table 1 Description of Datasets used**

| Dataset Name | Uncompr- essed Size (kB) | Size (mxt) And Rank | Remarks |
|---|---|---|---|
| TU Damstadt | 792 | 80x4902 Rank=28 | 80 appliances |
| UMass Microgrid | 5090 | 442x1440 Rank=389 | 442 individual houses |
| UMass Apartments | 476500 | 114x 522615 Rank=114 | 114 apartments |

## V. CONCLUSION

This paper presented a 3-stage data compression framework that is specially tailored for the need of high data compression in smart grids. The framework compresses the data in three different stages using different techniques. In the first stage, it exploits redundancy in the data using SVD for compression. The second stage applies normalization on resulting data whereas the third stage of Sparsity Encoding transforms the compressed output of second stage into its equivalent index value sparse matrix representation. Our simulation results show that our proposed data compression framework is very efficient for large non sparse datasets and small sparse datasets alike.

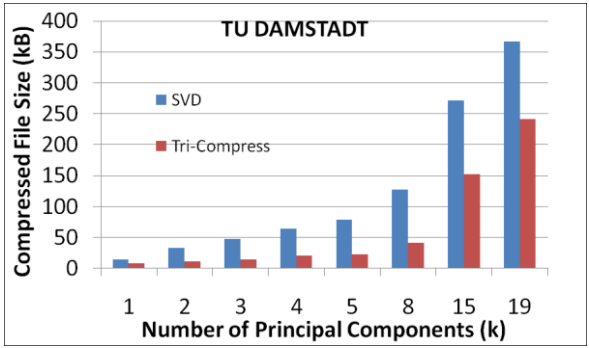 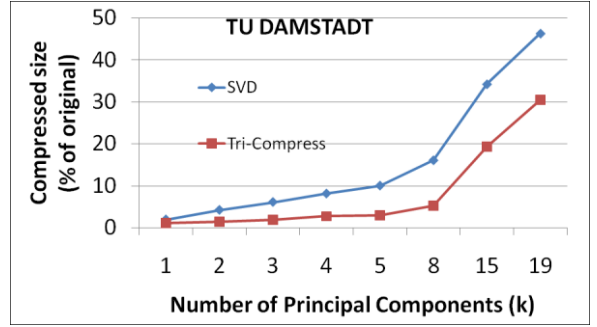

Figure 6. Performance on TU DAMSTADT dataset

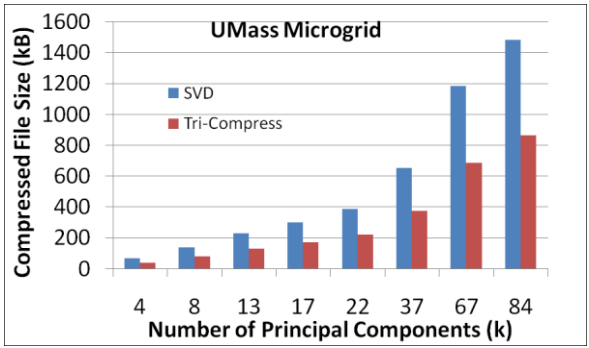 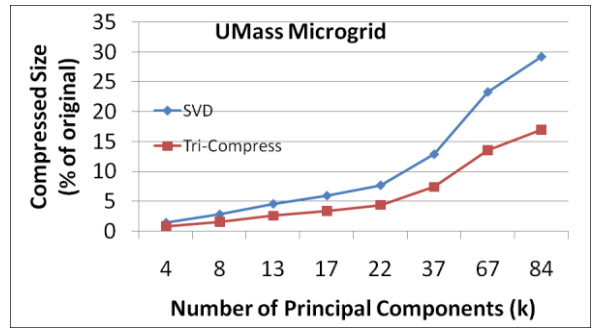

Figure 7. Performance on UMass Micro grid Dataset

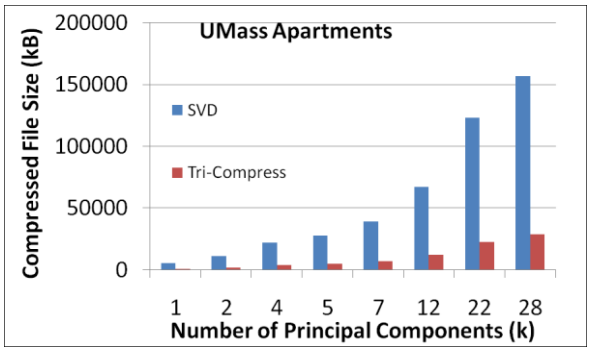 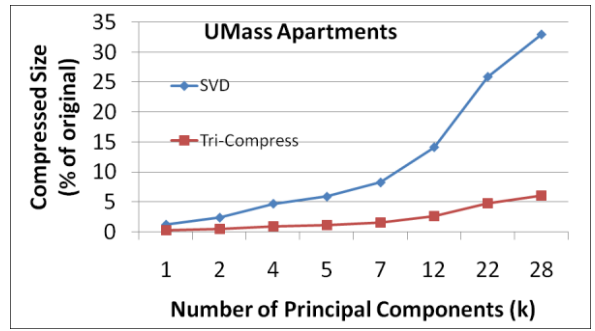

Figure 8. Performance on UMass Apartments Dataset

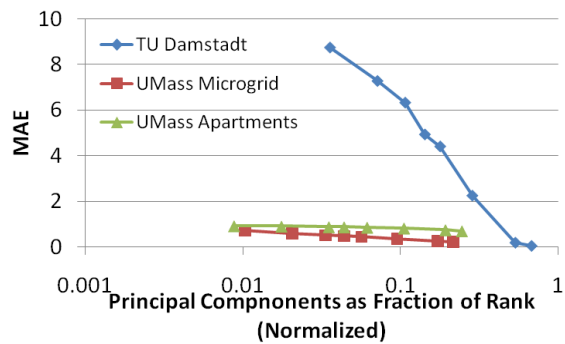

Figure 9. MAE Vs k as fraction of rank